# Grokking in the Ising Model

Karolina Hutchison and David Yevick
Department of Physics
University of Waterloo
Waterloo, ON N2L 3G1, Canada

## Abstract

Delayed generalization, termed grokking, in a machine learning calculation occurs when the increase in test accuracy is delayed relative to the training accuracy. This paper examines grokking in the context of a dense neural network trained to classify 2D Ising model configurations into 4 equally spaced energy regions in the presence of weight decay. Partially with the aid of novel PCA-based network layer analysis techniques, the observed behavior is interpreted as a transition from a connected network to a group of sparse subnetworks in which the number of active weights in each layer decreases monotonically with depth. This architecture reduces classification errors resulting from a multiplicity of paths. The final network layers, as in a convolutional neural network, sequentially identify global features of the input classes, which enables generalization to previously unseen patterns.

## Introduction

Recently, machine learning models were observed to generalize many epochs after initially overfitting the training dataset. While first reported in [1], this "grokking" behavior occurs in many neural network applications and architectures including modular arithmetic with transformers [1,2], image classification on the MNIST dataset using multi-layer perceptrons (MLPs) [3], sentiment analysis on the IMDb dataset using LSTMs [3], and natural language processing with transformers [4]. Non-neural models, such as classification and regression have also exhibited grokking behavior [5].

This paper demonstrates that grokking can occur when a dense neural network is applied to the Ising model in the presence of weight decay. The origin of grokking in this system results from the evolution of the network to an architecture in which successive layers contain a monotonically smaller number of active neurons. In such a network, the local properties of the spin distribution are processed by the initial layers while the global features, and hence the generalization capability, reside in the last layers with the smallest number of effective weights and neurons. The reduction of the neurons and weights in the final layers, however, requires many epochs resulting in the observed grokking.

The Ising model, viewed as a generator of images with maximum randomness subject to a given correlation among spins, provides an optimal framework in which to examine neural network evolution during training. Further, dense neural networks provide a far better framework than e.g.

CNN:s or physics-informed networks, in which to study grokking behavior qualitatively, as grokking is largely suppressed in the latter networks.

Grokking is affected by both the hyperparameters and the model architecture. Thus the authors of [1] found that the grokking delay, defined as the number of training steps required for a model to generalize after first reaching maximum training set accuracy, increases as the dataset size decreases, while increasing the weight decay or decreasing the initialization amplitude reduces the delay and hence improves generalization [3]. As well, the delay is affected by both the number of hidden layers and the number of neurons per layer [6]. Further, dropout can eliminate grokking entirely [7,8].

The training dynamics of a neural network can be analyzed from its hidden representation; in particular, its weight structure and subnetworks. Early studies of training attributed grokking to a decrease in the weight norms [3], but subsequent research indicated that this does not necessarily induce grokking but may correlate with it [8]. Further, both generalization [9] and grokking [10] can be associated with a transformation resembling a phase transition of the network structure in which a subset of neurons experience norm growth or norm decay. The weight matrices accordingly transition to a low-rank configuration resulting in a sparse subnetwork [11]. The resulting reduction of the local complexity of the network's input-output mapping results in simplified decision boundaries in the input space [6]. The activation sparsity, e.g. the fraction of neurons in a **ReLU** hidden layer that are inactive, also tends to increase during training when grokking occurs, reaching a plateau in the region where the training accuracy is 100% but the test accuracy is low [8]. This behaviour has also been studied in transformers, where the **ReLU** multi-layer perceptron (MLP) exhibits sparsity after training [12]. These effects can also be quantified from the evolution of the gradient vectors generated during optimization, which is conveniently visualized through dimensionality reduction methods such as principal component analysis (PCA) [13,14].

**Computational Results:**

To observe grokking in the 2D Ising model, a training dataset consisting of 4000 8x8 lattices with uniformly distributed energies was constructed. Each lattice was initialized with a different random spin configuration to ensure that the samples are uncorrelated. The Metropolis algorithm with a decreasing temperature schedule was then applied to obtain configurations spanning a range of energies, defined by the standard formula $E = -\sum_{\langle ij \rangle} s_i s_j$, in which the sum is evaluated over all nearest-neighbour pairs with periodic boundary conditions and $s = \pm 1$. For an 8x8 lattice, the theoretical minimum and maximum energies are -128 and +128 respectively. However, the dataset only covers a subset of this range such that randomly initialized configurations possess near-zero energies, while the ground state energy is -128. The dataset is then sorted into four evenly spaced energy intervals, $\Delta E_n \equiv \Delta E_1, \dots, \Delta E_4$, each containing 1000 data records. The categorical label employed in our neural network examples is given by the energy index $n = 1,2, . . 4$. The energy ranges for these intervals are shown in Table 1, where the data samples were uniformly distributed in energy within each of the ranges. Representative samples from each of these intervals are displayed in Figure 1. While these energy intervals of course could be defined differently or be based on the degree of disorder, additional calculations demonstrated that the results were effectively identical when changing the number or ranges of the energy intervals or when the intervals are defined by magnetization instead of energy. This behavior is expected since the mechanics of grokking is independent of the selection of output categories.

| Energy category | Energies |
|---|---|
| $\Delta E_1$ | -120 to -92 |
| $\Delta E_2$ | -88 to -60 |
| $\Delta E_3$ | -56 to -28 |
| $\Delta E_4$ | -24 to 4 |

*Table* 1. *The four energy intervals of the training dataset.*

*Figure 1. Examples of training data. The rows are the four energy categories*

Two test datasets were generated. The first, "random spin test set," which is employed below unless indicated, was obtained in an identical manner to the training dataset, while the samples in the second, "$n$-inverted spin test set," were produced by reversing $n$ randomly selected spins in each lattice of the training set. In this case, the energy of each lattice was recalculated, and the test labels were reassigned accordingly. In the subsequent calculations, $n = 5$ unless otherwise specified. As evident in Table 2, the number of states in each $\Delta E_i$ bin then differs from the uniform distribution of the training data with $\approx 52\%$ of lattices changing energy category. Nearly all of the $\Delta E_1$ samples were displaced into the $\Delta E_2$ bin while the majority of the $\Delta E_2$ samples entered the $\Delta E_3$ bin and most of the $\Delta E_3$ samples and nearly all of the $\Delta E_4$ samples remained in their original bin. In contrast when $n = 2$, after spin inversion $\approx 80\%$ of the lattices remain in the same energy category. Accordingly, the dependence of grokking on the similarity between the training and test data can be quantified by comparing the response of the neural network to these two data sets.

| Energy category | Number of lattices |
| --- | --- |
| $\Delta E_1$ | 40 |
| $\Delta E_2$ | 1214 |
| $\Delta E_3$ | 1409 |
| $\Delta E_4$ | 1337 |

*Table 2.* The number of lattices in each energy category after randomly inverting 5 spins to generate the test set.

Once constructed, the training and test sets were applied to a multi-layer perceptron (MLP) neural network in which the output class label was given by the energy category, $\Delta E_i$. Unless otherwise stated, the network employed four 48-neuron hidden layers with **ReLU** activation functions, cross-entropy loss, a weight decay of $5 \times 10^{-4}$, and a learning rate of $5 \times 10^{-4}$. A **softmax** activation function is applied in the final network layer modified by the introduction of an inverse temperature parameter, $\beta$, into the **softmax** function according to $P_i = \exp(\beta z_i)/\sum_{j=1}^{4} \exp(\beta z_j)$, where $z_i$ denotes the $i$-th logit from the output layer. By default, $\beta = 1$. To simplify the analysis, the bias terms in the ReLU neurons were omitted and the batch size was set to the dataset size, so that gradient updates were only performed once per epoch. The initial weights were sampled from a scaled uniform distribution, $c \times U$, where $U$ denotes the PyTorch uniform distribution and $c$ denotes the initialization amplitude (default $c = 7$). The weight initializations and decay factor were selected in order to enable grokking, as noted in [3]. As evident from the theoretical discussion in the following section, the large initial initialization amplitudes delay the transition to a sparse network with high test accuracy while the decay factor suppresses superfluous weights.

With these values, the test accuracy asymptotically approached $\approx 0.83$ and $\approx 0.79$ for the random spin and $n = 5$ inverted spin test sets, respectively, after an extended period of overfitting. As expected, the greatest source of error originated from predictions that were incorrect by one energy interval. To determine if the neural network in the overfitting phase preferentially associates the energy of a given inverted spin lattice with the energy of the original non-inverted lattice, an $n = 2$ calculation was performed for which the asymptotic test accuracy was approximately $0.85$. The test accuracy was only $0.55$ during overfitting. An analysis of the $4 \times 4$ confusion matrix for the test set after 5000 epochs in Figure 2 indicates that while overfitting, the neural network was roughly equally likely to misclassify samples into higher and lower energy bins. This implies that even though the training and test sets are nearly identical, the overfitting neural network does not at first exploit this similarity, so that its predictions are initially meaningless and uncorrelated with the underlying physics, almost independent of the value of $n$.

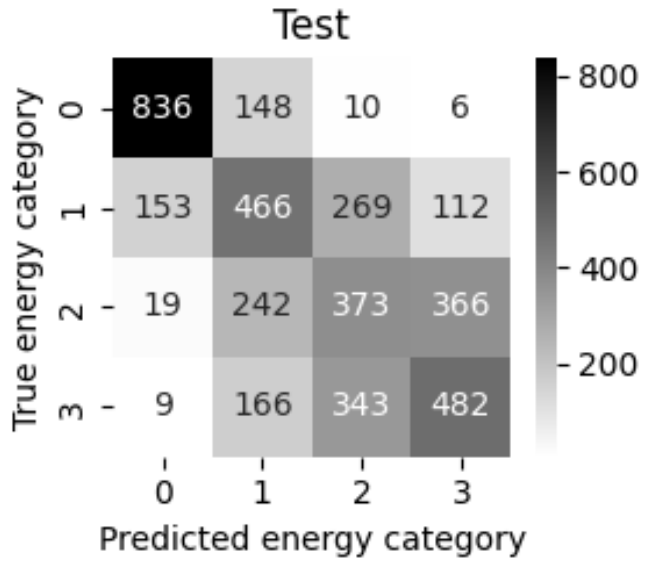

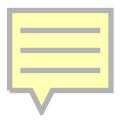

*Figure 2. The test set confusion matrix after 5000 epochs for the random spin test set.*

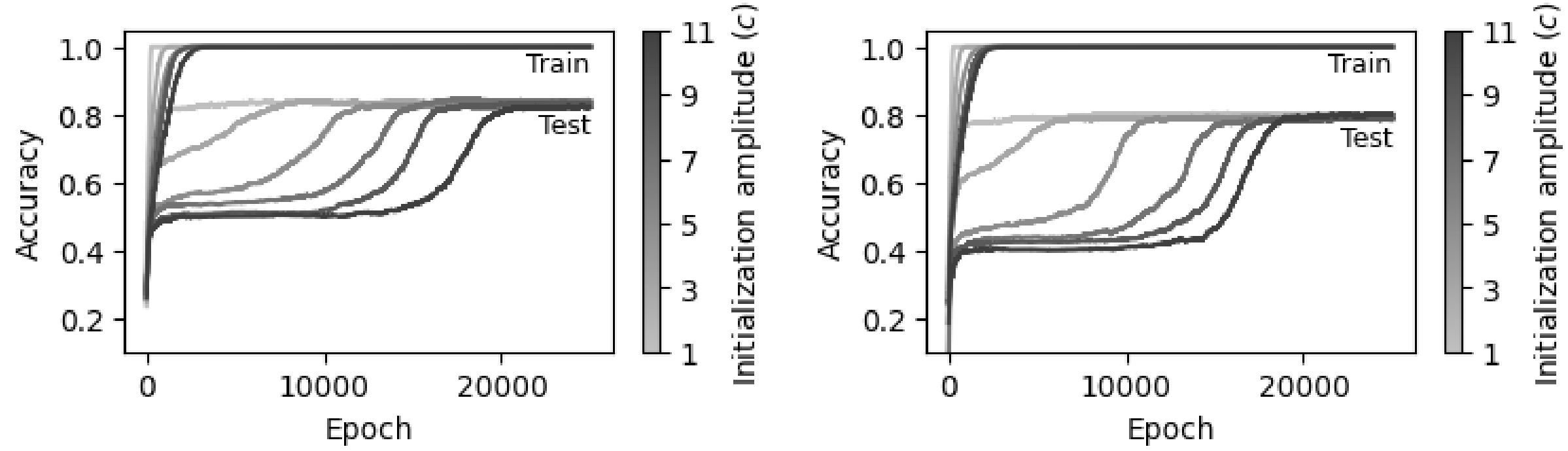

*Figure 3. The training and test accuracy curves for varying initialization amplitudes, for (a) the random test dataset and (b) the 5-inverted spin test dataset.*

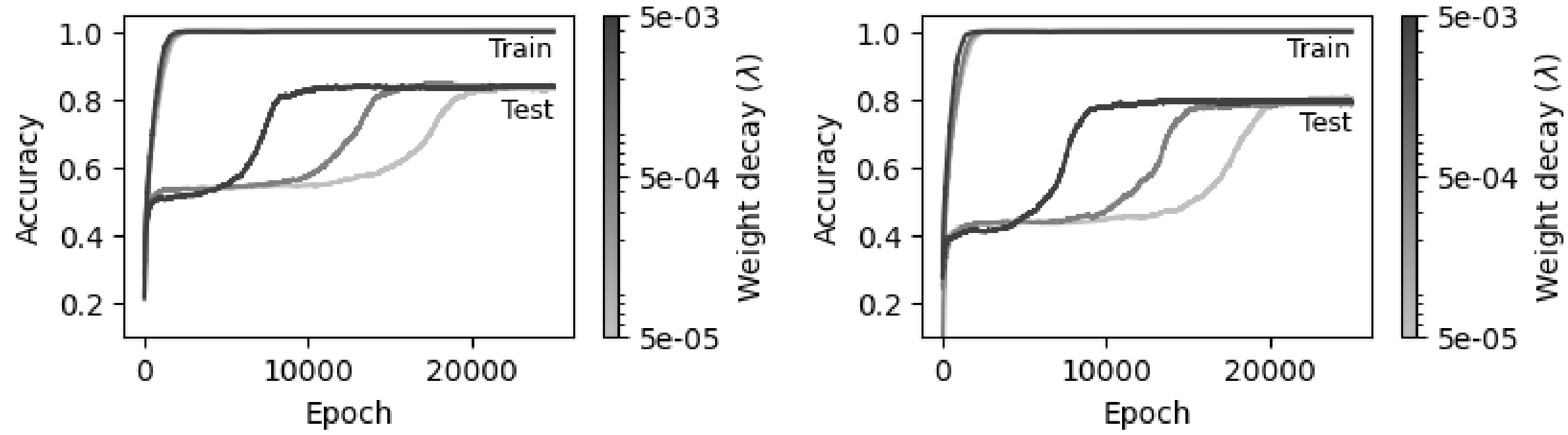

*Figure 4. The training and test accuracy curves for varying values of weight decay, for (a) the random test dataset and (b) the 5-inverted spin test dataset.*

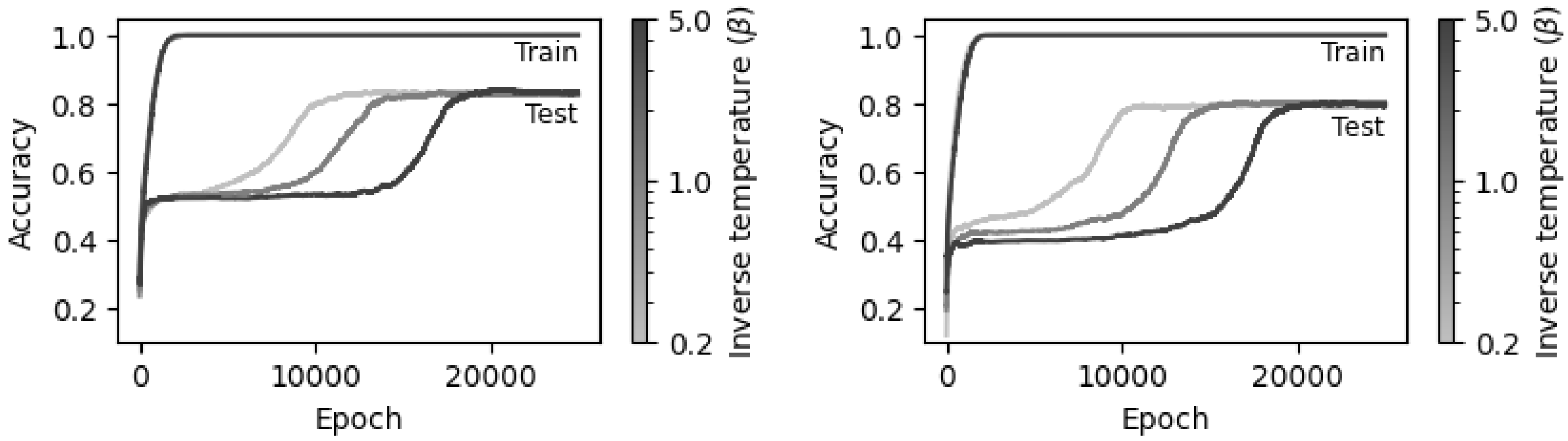

*Figure 5. The training and test accuracy curves for varying values of inverse temperature in the* ***softmax*** *function, for (a) the random test dataset and (b) the 5-inverted spin test dataset.*

As in [3], the grokking delay was found to increase with the initialization amplitude, $c$, and decrease with the weight decay parameter, $\lambda$, as evident from Figure 3 and Figure 4, which show the training and test accuracy as functions of $c$ and $\lambda$, respectively (when $\lambda = 0$, the model does not generalize as the test accuracy remains constant at $\approx 0.55$ and $\approx 0.46$ for the random and $n = 5$ lattices respectively). However, grokking is suppressed when the default initialization was employed as seen from Figure 3. These results demonstrate that the previously reported dependence of grokking on initialization and weight decay [3] additionally applies to the Ising model. Figure 5 demonstrates that increasing the temperature in the **softmax** function (e.g. decreasing $\beta$) reduces grokking. This is consistent with prior work, which found that higher temperatures lead to faster convergence [15]. Furthermore, the grokking delay was found to increase with the number of hidden layers and the hidden layer size.

To understand better the source of grokking in the Ising model, the global neural network structure can be visualized during training. Figure 6-Figure 9 display the interactions among the four hidden neural network layers where the widths of the lines represent the squared values of the positive and negative weights rescaled to fit within $[0, 1]$. The associated accuracy curves are shown to the left of each figure, where the circular markers indicate the epoch at which the network diagram was extracted.

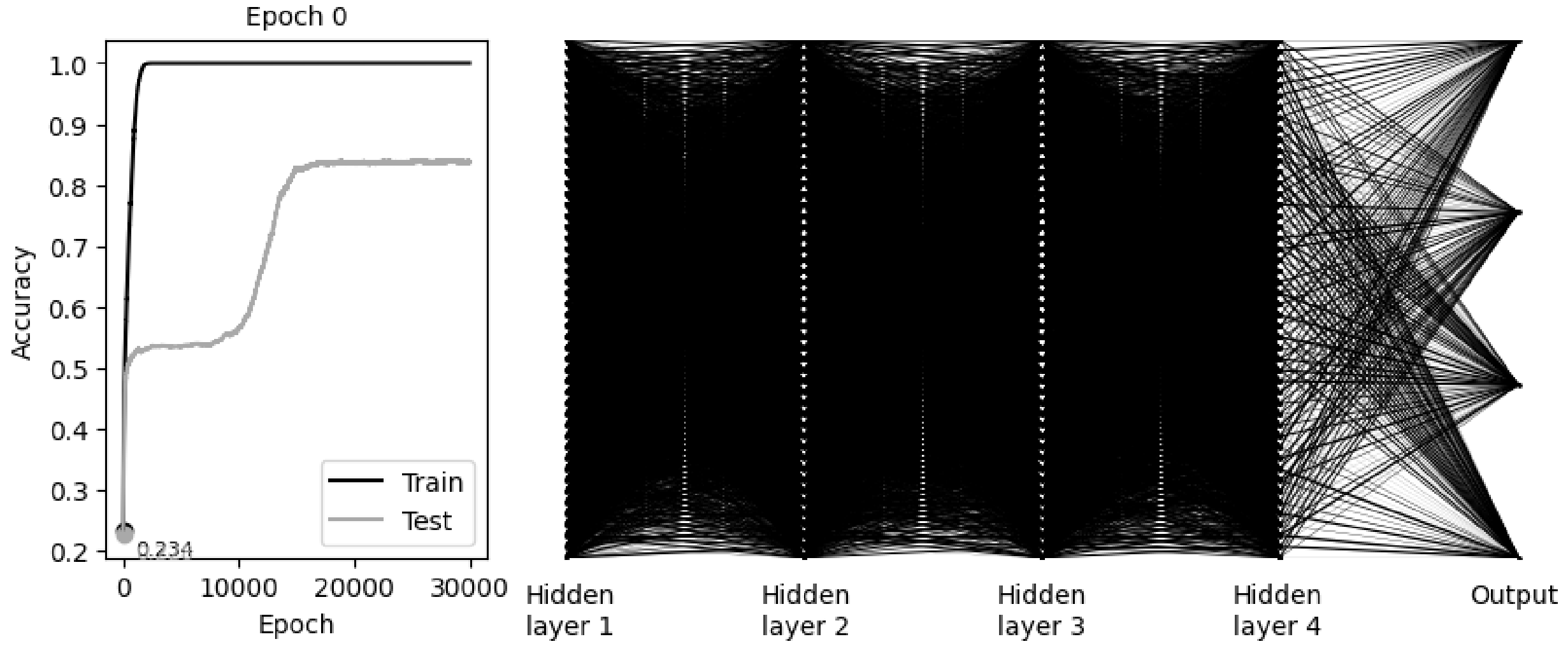

*Figure 6. The weights at initialization (0 epochs).*

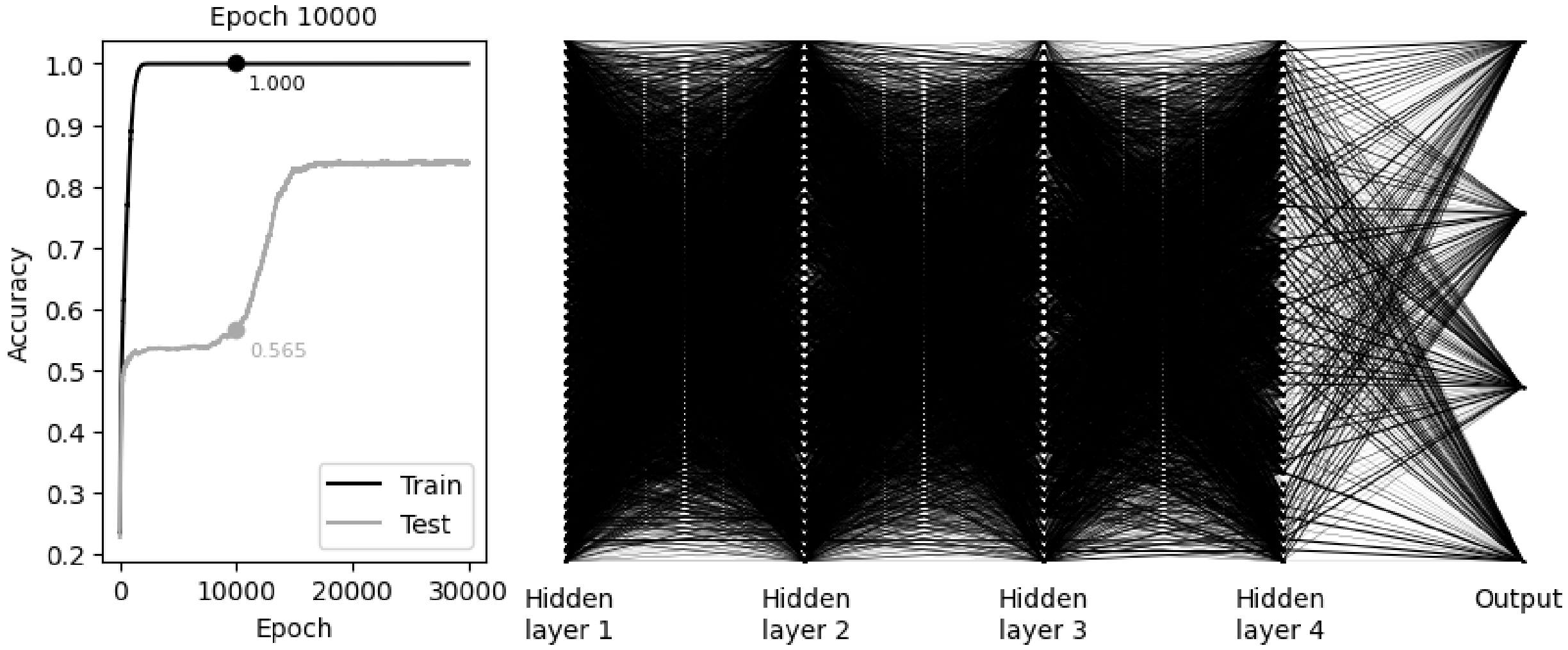

*Figure 7. The weights near the end of the overfitting region (after 10000 epochs).*

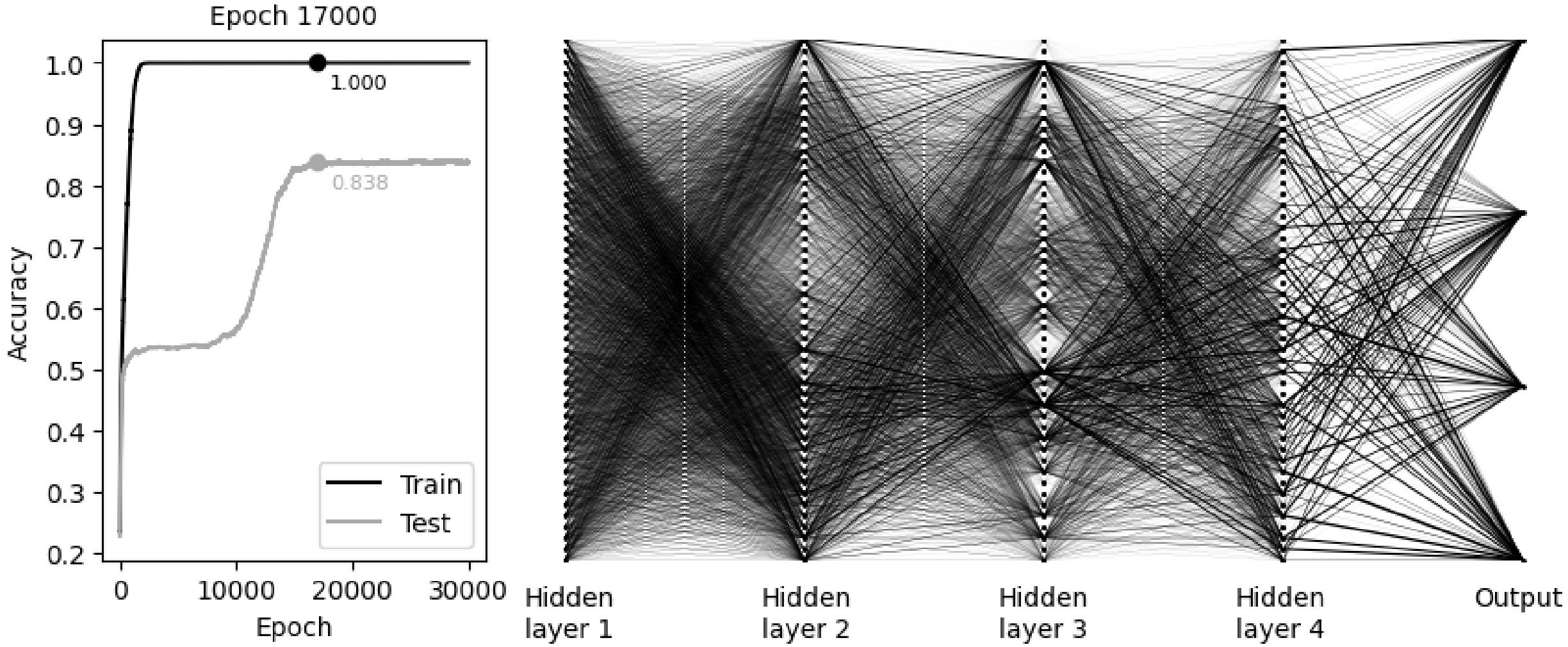

*Figure 8. The weights shortly after grokking (after 17000 epochs).*

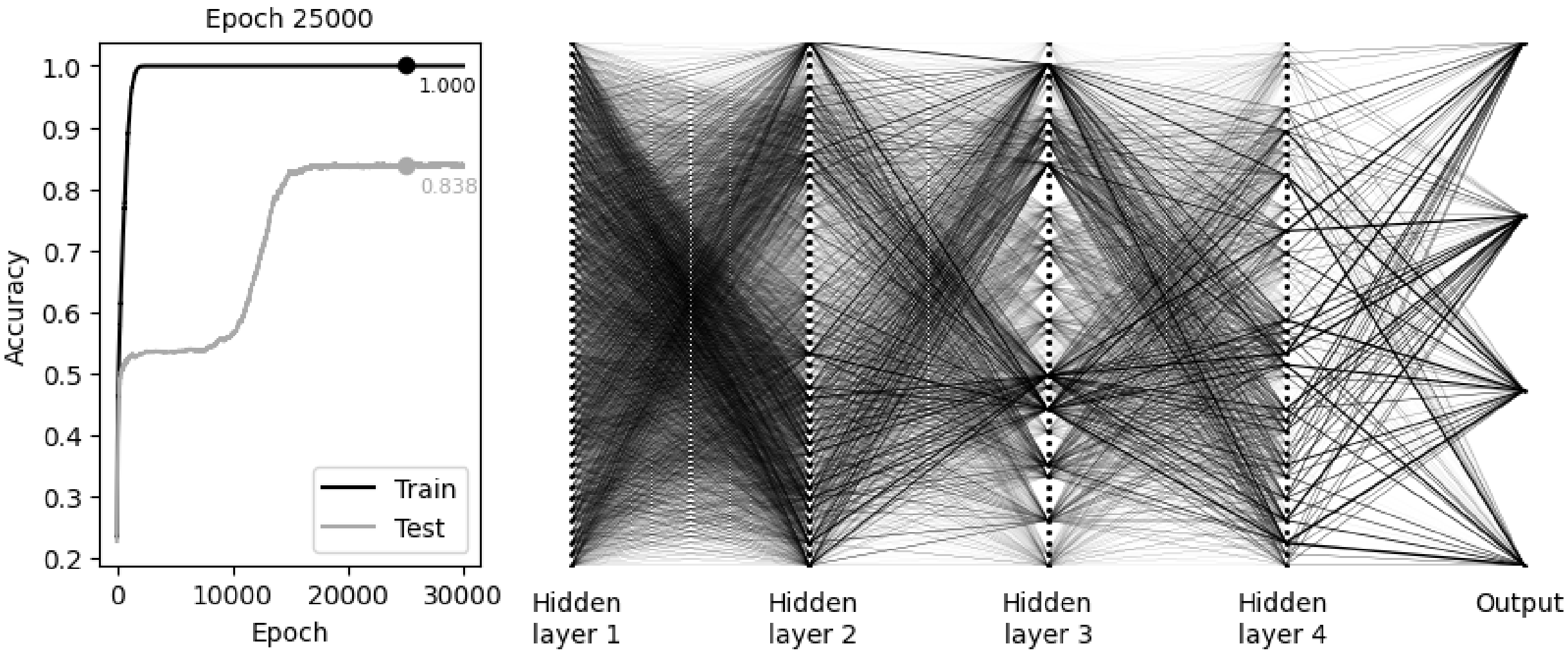

*Figure 9. The weights long after grokking (after 25000 epochs).*

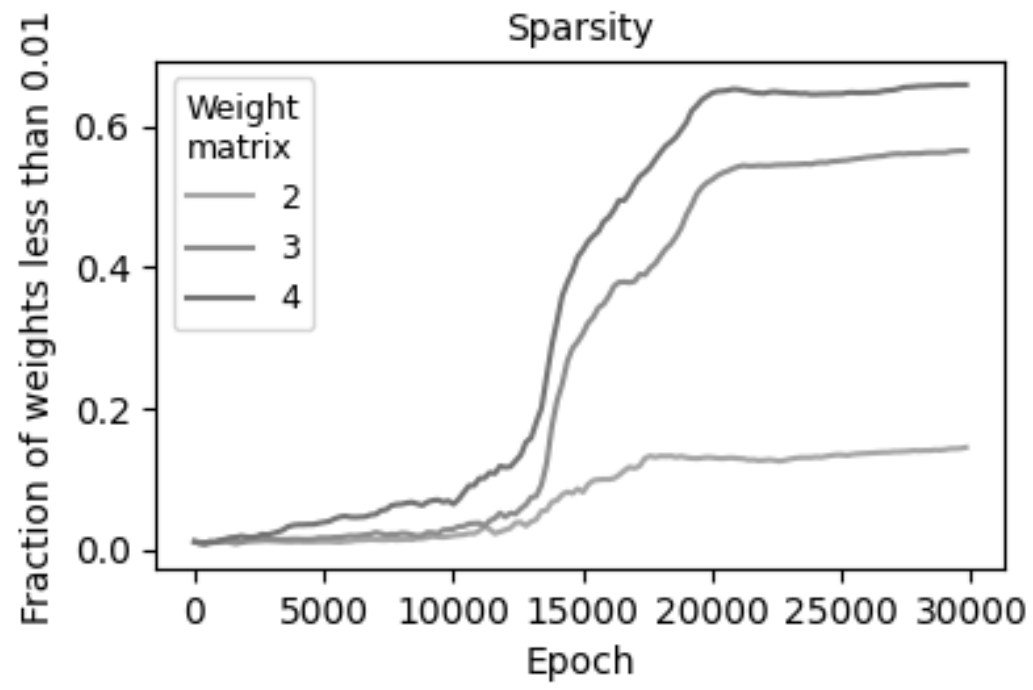

*Figure 10. The fraction of weights in each of the* $48 \times 48$ *hidden weight matrices with magnitude <0.01 in Figure 6-Figure 9.*

During grokking, between 10000 and 17000 epochs, a substantial fraction of the neurons become inactive while many weights approach zero, resulting in a transition to a sparse network, as first reported in [11]. This behavior is illustrated in Figure 10 where the fraction of weights between hidden layers 1 and 2, 2 and 3 and 3 and 4 in Figure 6-Figure 9, all of which possess an identical number of neurons, with amplitudes less than 0.01 is graphed as a function of epoch number. This figure clearly indicates that the network transitions to an architecture with a sparsity that increases with layer depth as the network groks. After the neural network generalizes, however, its structure changes only slightly with epoch number. Similar results are obtained if the MSE with one-hot categorical labels is employed in place of the cross-entropy loss function, or if the **tanh** activation function is substituted for **ReLU**. That the increase in sparsity is associated with a combination of weight decay and initialization amplitude is further evident from Figure 11 which is generated from a standard dense network calculation with the Pytorch default initialization and zero weight decay. In this case, the sparsity does not change with epoch number and consequently grokking is absent.

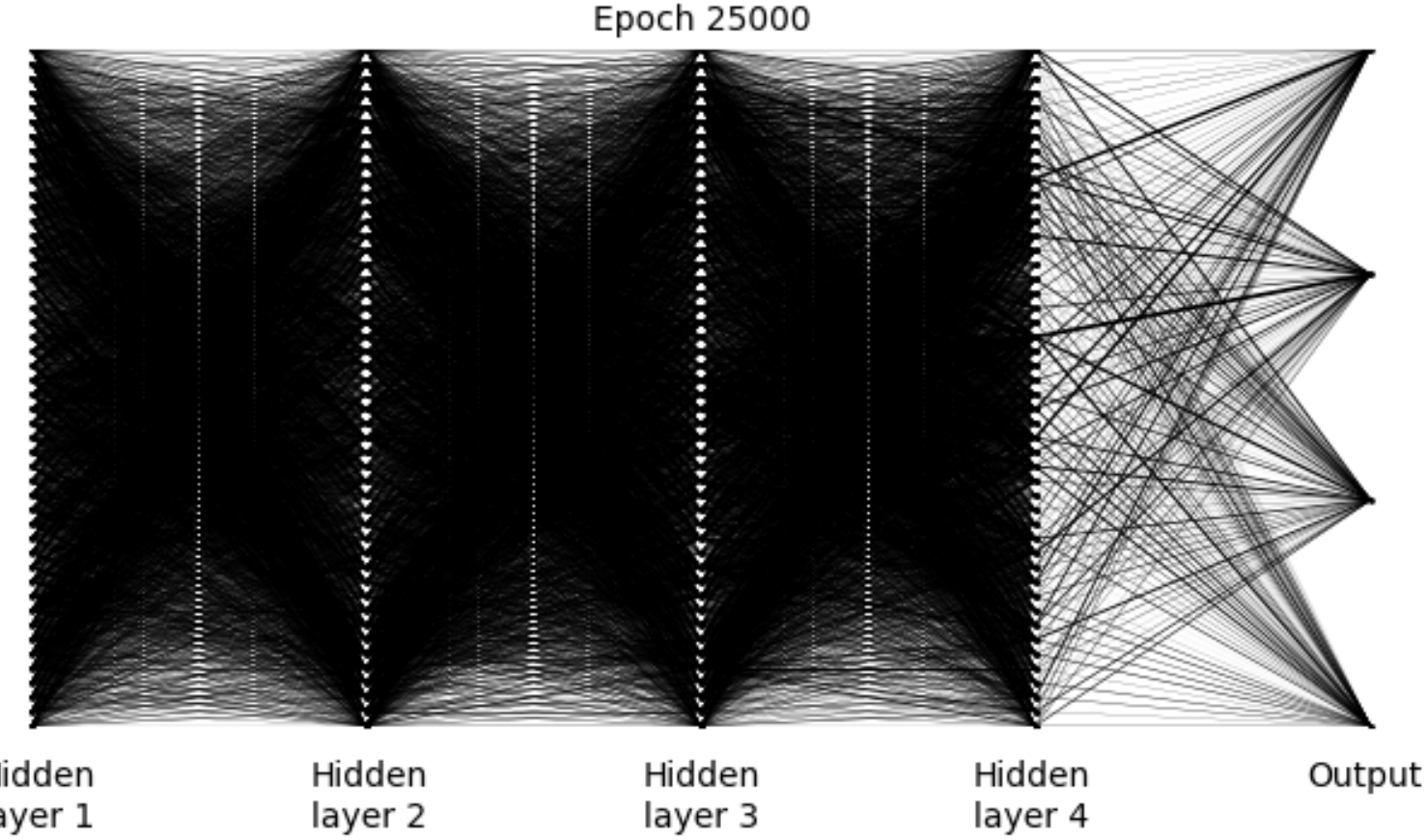

*Figure 11. The neural network weights after 25000 epochs in the absence of weight decay for the default initialization.*

The mechanism responsible for grokking can be visualized by first observing that a feedforward neural network implements a recursive transformation of input data vectors $\boldsymbol{x}$ given by

$$f_{\theta}(\boldsymbol{x}) = h(\boldsymbol{A}_{l}\mathrm{h}(\boldsymbol{A}_{\boldsymbol{l-1}}\mathrm{h}(\dots) + \boldsymbol{b}_{\boldsymbol{l-1}}) + \boldsymbol{b}_{\boldsymbol{l}}) \quad (1)$$

Here $\boldsymbol{A}_i$ and $\boldsymbol{b}_i$ are weight matrices and bias vectors, respectively, and the final, $l$:th, term in the recursion is $h(\boldsymbol{A}_{\boldsymbol{1}}\boldsymbol{x} + \boldsymbol{b}_{\boldsymbol{1}})$. If the $h(\boldsymbol{x})$ are **ReLU** activation functions, the input space is accordingly partitioned into piecewise affine spline operators according to

$$f_{\theta}(\boldsymbol{x}) = \sum_{\omega \in \Omega} (\boldsymbol{A}_{\omega}\boldsymbol{x} + \boldsymbol{b}_{\omega})\mathbf{I}_{(x \in \omega)} \quad (2)$$

where $\boldsymbol{A}_{\omega}$ and $\boldsymbol{b}_{\omega}$ are new local weight matrix and bias array parameters while $\mathbf{I}_{(x\in\omega)}$ is an indicator function that is unity in a "linear region" $\omega \in \Omega$ and zero elsewhere such that the nonoverlapping regions $\Omega$ span the neural network's input space. The density of these linear regions in the immediate neighbourhood of a point $\boldsymbol{x}$ in the input space provides an estimate of the local complexity at this position. Specialized visualization tools additionally demonstrate that during a grokking transition mediated by weight decay (regularization) linear regions migrate toward the decision boundary resulting in a large curvature transverse to the boundary surface while the complexity of the remaining regions decreases [6].

To extend the above framework, note that the sparsity of the layers in Figure 10 increases with layer depth. This suggests an analogy with a CNN in which the initial network layers contain many active neurons and accordingly capture the fine structure of the input data while the final, sparse, layers process averaged, global, features of the data. Similarly, Eq.(1) implies that the initial layer, which possesses many active neurons and weights that directly process the input data, affects the properties of the smallest linear regions (since the output space of the first layer is divided repeatedly by the subsequent activation functions). In contrast, the functions $h(x)$ associated with the final layer, which contains relatively few participating neurons and weights. are instead applied to the largest consolidated groups of input variables. These layers therefore promote generalization by modifying extended regions of the input space. Before grokking, the large number of neurons present in a dense network rapidly adjust to memorize the input data. However because of the large number of channels present in the final layer, slight deviations from the training set generate unpredictable false predictions. Once, however many neurons and weights are suppressed in the final layers the resulting network can reliably generalize to the test set.

A related explanation of the above results follows from the shape of the decision boundaries between classes which form hypersurfaces in the high-dimensional space of the input variables. Because of this dimensionality, the probability that a given test data record, viewed as a point in the input space, is positioned close to one or more of these hypersurfaces is large. Hence the local properties of the decision boundary significantly affect the test accuracy. In the initial memorization phase, the decision boundary surfaces are corrugated by the action of the large number of participating neurons, indicating that the dense network has overfit the training data and therefore decreased the test set classification accuracy. The subsequent reduction in the complexity of the final network layers, however, smooths the decision boundaries to reflect the average properties of the training set data, enabling generalization.

Next, a dropout layer parameterized by the dropout rate $p$, defined as the probability that a given unit in the output of each hidden layer is set to zero during a forward pass, was added after each of the four hidden layers. To obtain optimal results, the hidden layer size was additionally increased to 144, although the results for other sizes were qualitatively identical. The training and test accuracies and corresponding losses, averaged over 4 runs, are shown for different dropout rates in Figure 12 and Figure 13, with a weight decay factor of $\lambda = 5 \times 10^{-4}$. While dropout was applied during training, the training accuracy in these curves was calculated with the full neural network.

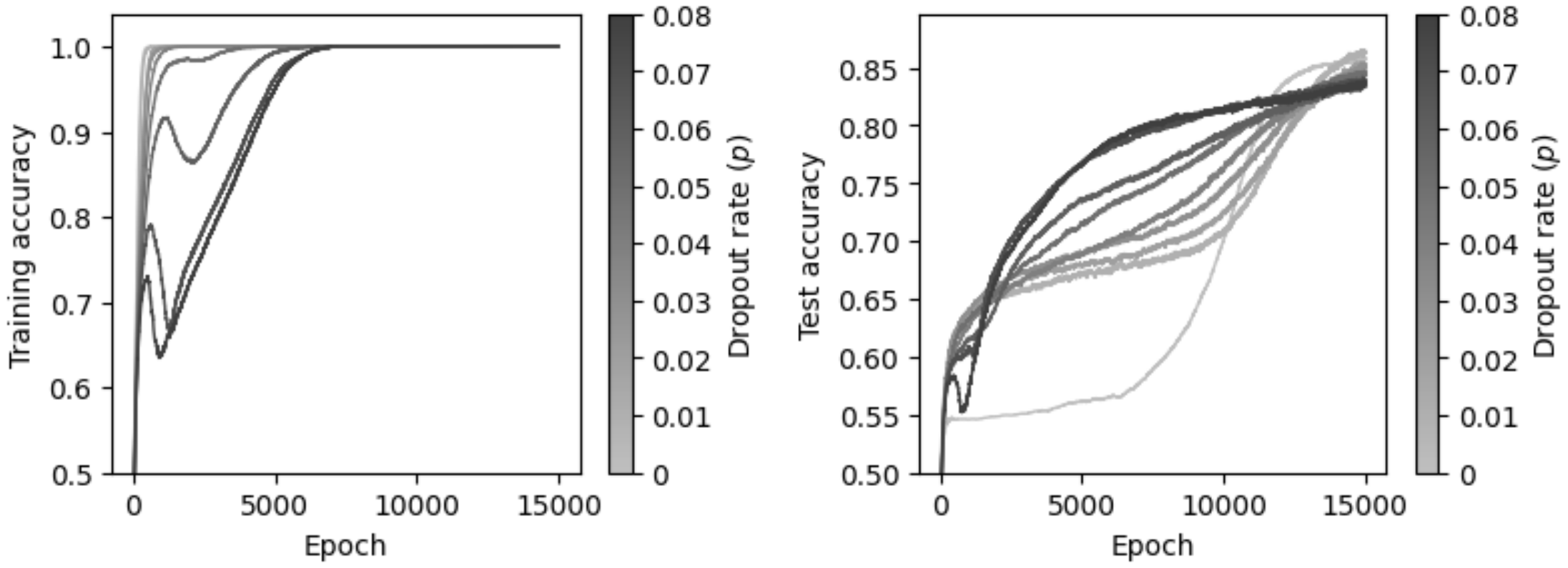

*Figure 12. The accuracy curves for the (a) training and (b) test datasets for different dropout rates.*

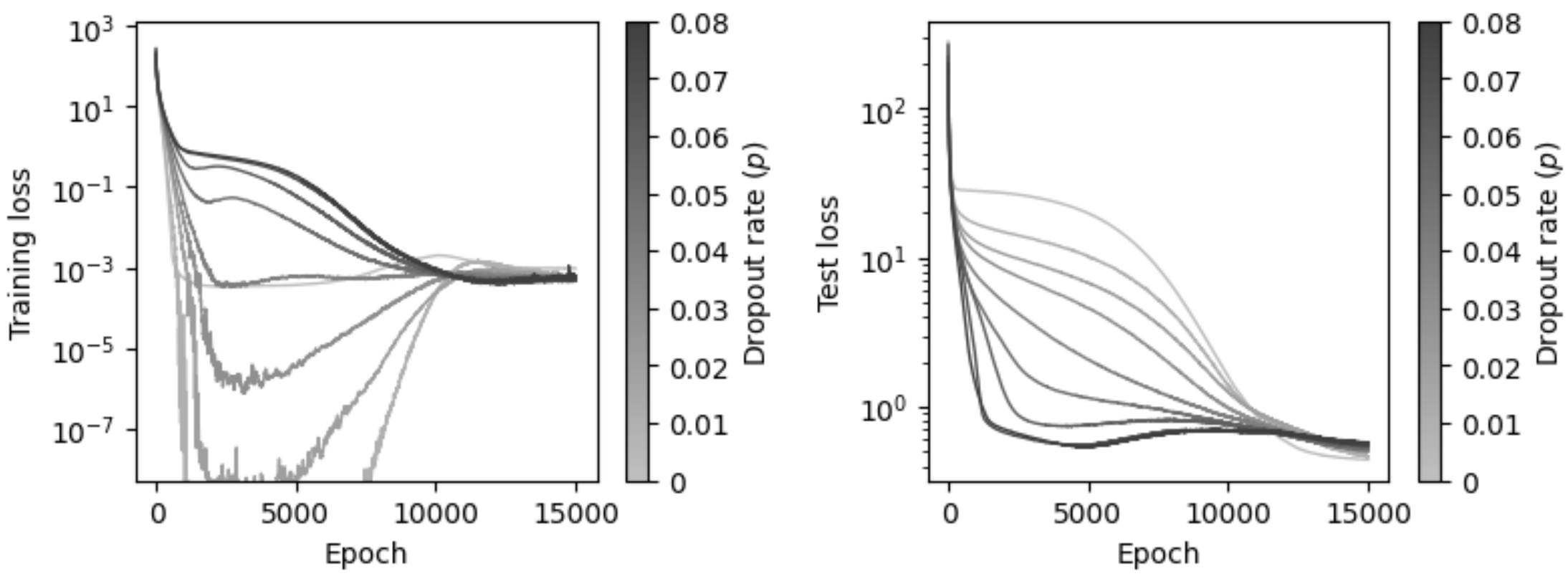

*Figure 13. The loss curves for the (a) training and (b) test datasets for different dropout rates.*

Dropout rates as small as $p = 0.04$ are sufficient to accelerate generalization and hence suppress grokking, as expected from [7,8]. For $p >\approx 0.05$, the training accuracy initially becomes smaller and then increases. Both the training and test accuracies then increase roughly simultaneously, while an analysis similar to that of Figure 6-Figure 9 indicates that the number of significant network neurons remains roughly constant. This is expected since the random elimination of neurons precludes the formation of a stable subnetwork. These observations are only slightly affected by the magnitude of the weight decay parameter.

Further insight into grokking can be obtained by viewing the lowest-order PCA components of the gradients of the loss with respect to the parameters of the five weight matrices present in the neural network for each data sample. These quantities are displayed as dots in Figure 14.

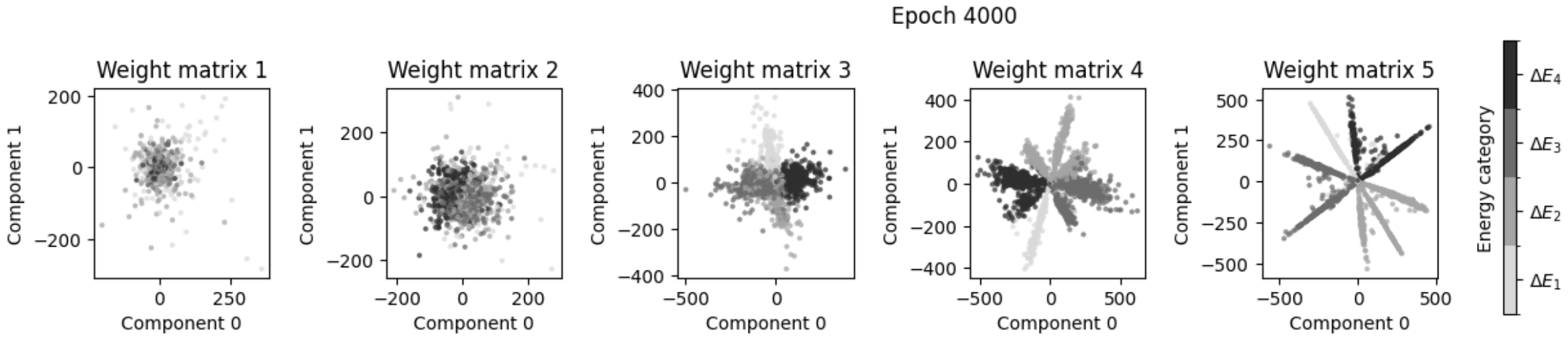

*Figure 14. A sample of the PCA-reduced gradients of the test losses, at epoch 4000 where the neural network is overfitting.*

The lines of dots that become more pronounced in the later network layers are formed by groups of misclassified samples, while the gradients of the correctly classified samples are situated near the center of the diagrams.

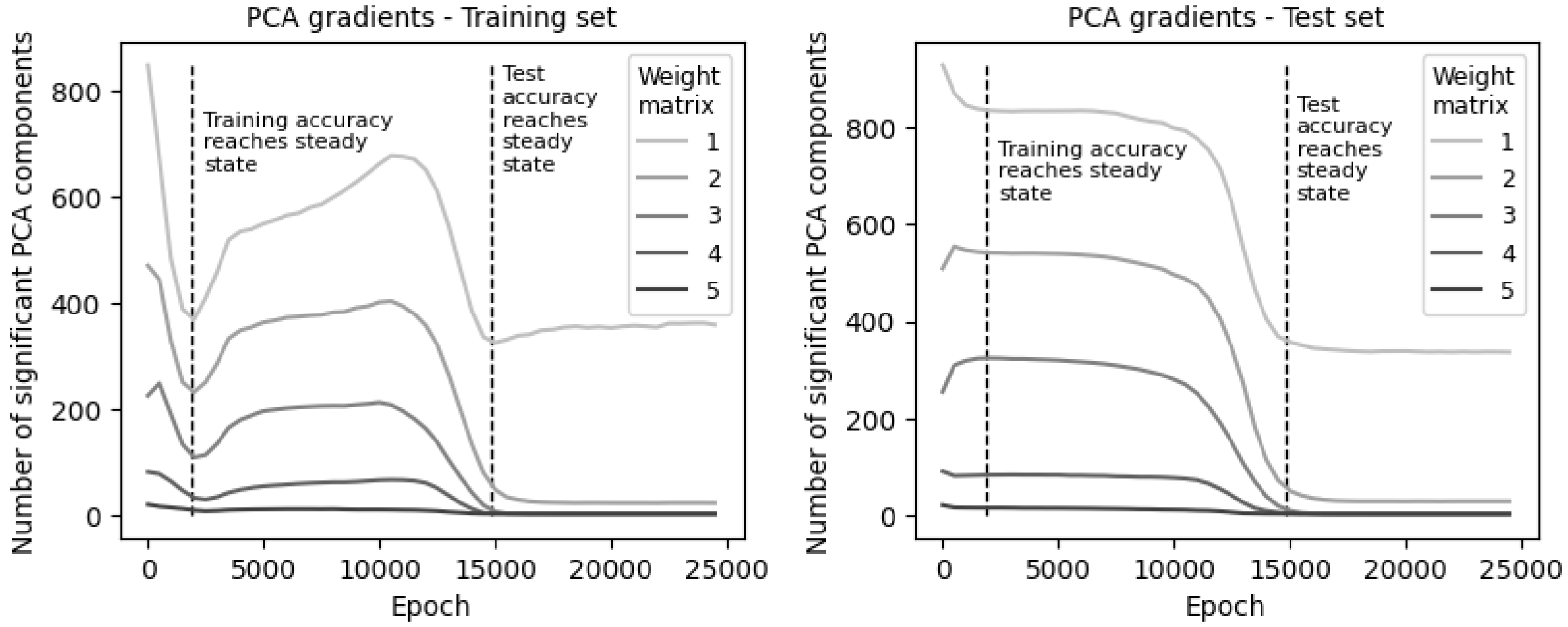

*Figure 15. The number of PCA eigenvectors required to account for 90% of the total variance.*

A further, apparently novel result is obtained by considering the number of PCA components required to account for 90% of the variance in the data. This quantity, calculated every 500 epochs and averaged over 5 consecutive calculations, is displayed in Figure 15, where the lines designating the points at which the training and test accuracy approximate their asymptotic values are estimated from the previous graphs. These curves reflect the degree of uncertainty in the optimization step, since if many PCA dimensions are significant, the gradient (which is a stochastically varying quantity dependent on the minibatch content) can be oriented along and therefore sample a far larger set of directions. Evidently, the loss landscape of the training dataset increases in complexity during grokking resulting in a more effective minimization of the loss function and an accompanying simplification of the landscape.

PCA can be used to visualize not only the gradients but also the output activations of each hidden layer for all samples in the dataset at each training stage. The neurons of each hidden layer yield a 48-dimensional vector for each sample, which is then projected onto the two lowest-order PCA components in Figure 16-Figure 18. While these figures were generated with the random test set, similar results were obtained with the $n = 5$ dataset.

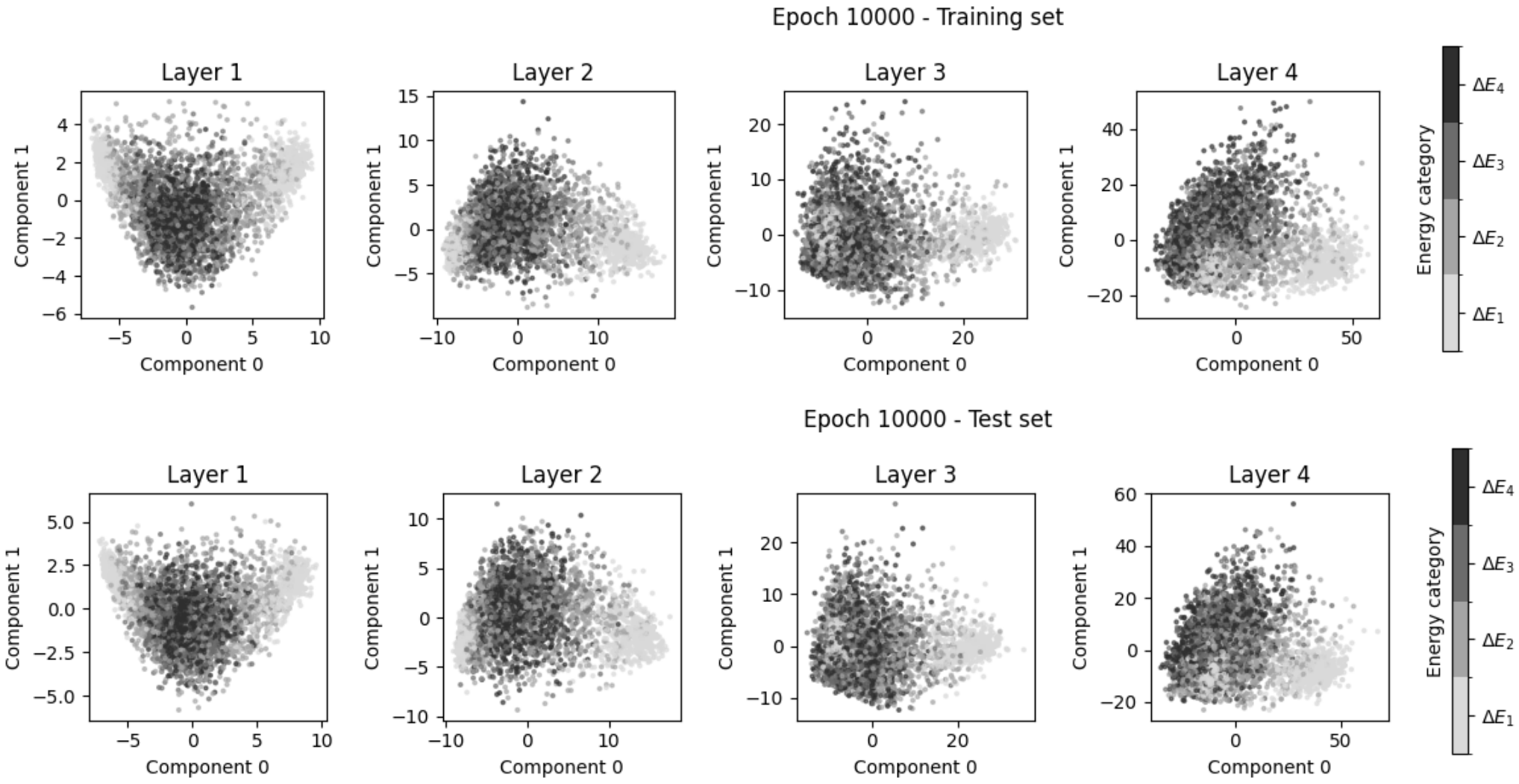

*Figure 16. PCA projections of the hidden layer representations for (a) the training set and (b) the test set when the network has reached 100% training accuracy, and the test accuracy is beginning to increase.*

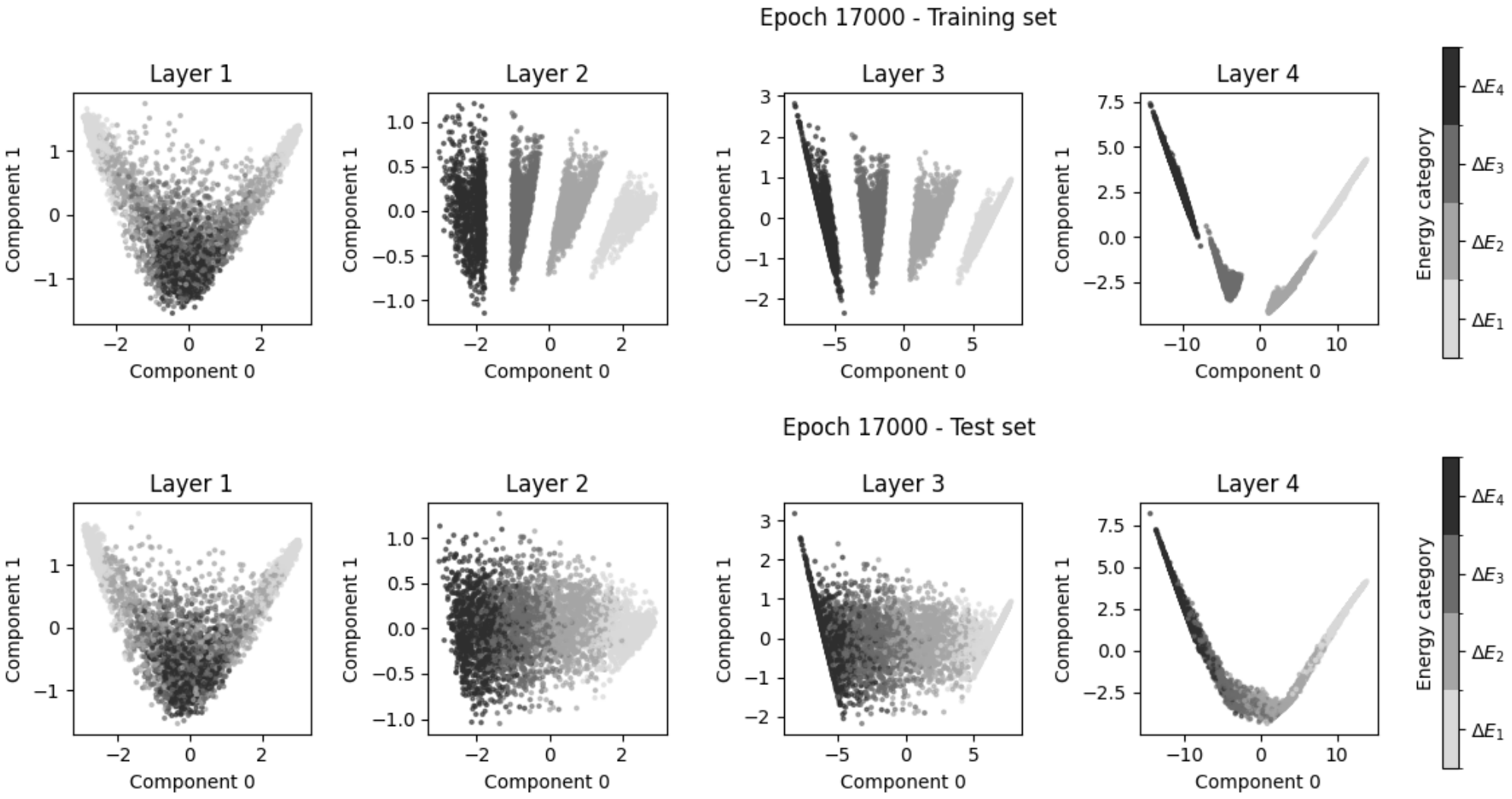

*Figure 17. The same as Figure 16, except for epoch 17000, just after the test accuracy has attained its asymptotic value.*

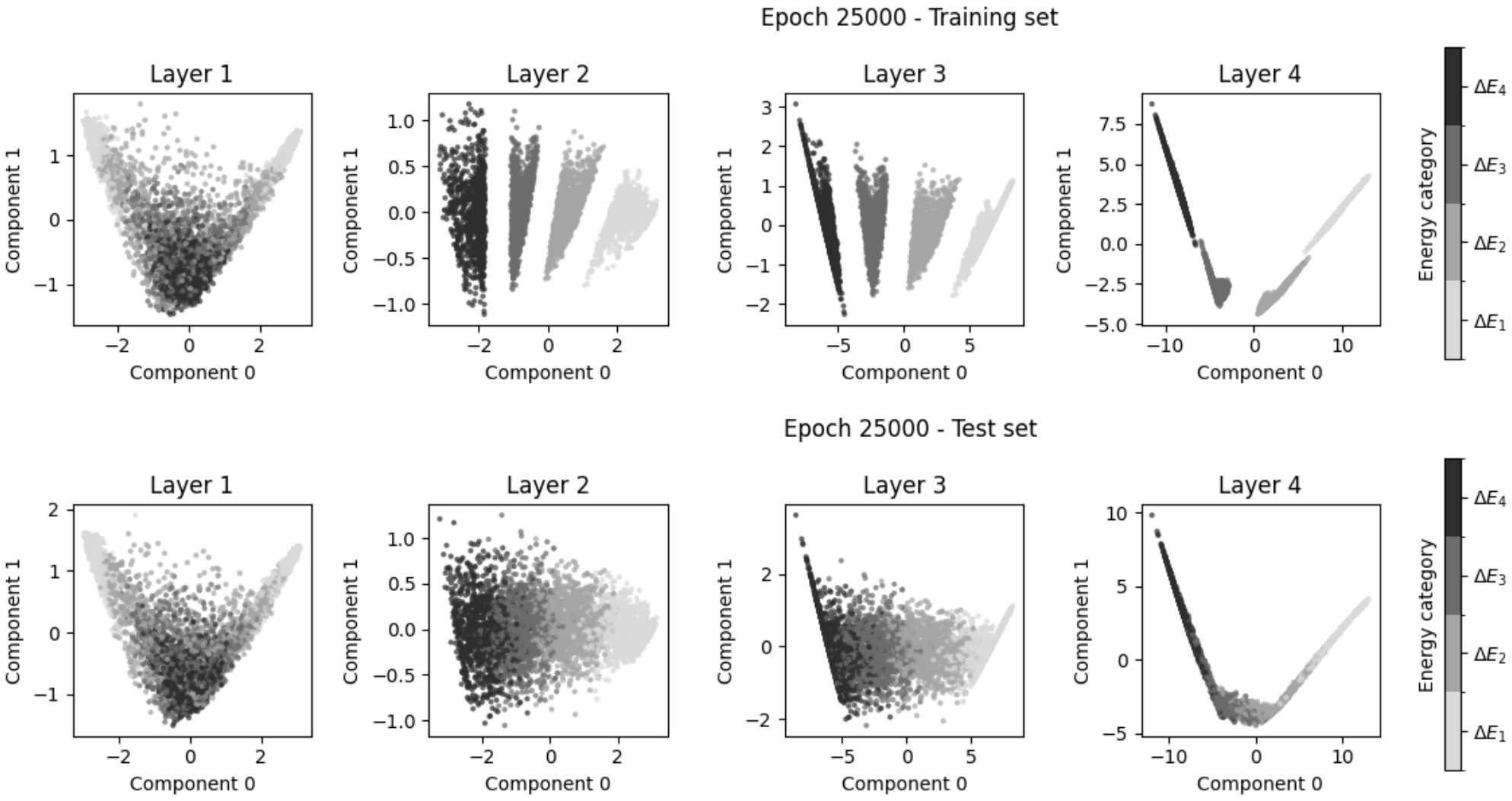

*Figure 18. The same as Figure 16, except for epoch 25000.*

Figure 16-Figure 18 demonstrate that even after the training accuracy reaches 100% the internal representations of the dataset are unstructured; at this point, the network has not yet conclusively identified the input distributions. As grokking begins, however, a transition occurs that organizes the activation vectors into distinct clusters corresponding to energy categories. The network is then able to recall the features of each input record, in agreement with prior work that associated grokking with a simplification of the decision boundaries [6]. At least in the ReLU case, once the activation patterns become more structured, many neurons become inactive and therefore experience norm decay. The network can then accurately identify the mapping between the input data and the classes and resolve the perturbed input patterns present in the test set.

Additional insight into the grokking mechanism can be obtained by examining the output "probability" distribution constructed from the output values of the **softmax** layer. In particular, the mean of the difference, $\overline{\Delta P} = \overline{P_{max} - P_{second}}$, between the largest and second-largest predicted "probabilities", provides a heuristic but easily evaluated metric for the prediction confidence, or alternatively, the output entropy. As evident from Figure 19(a), $\overline{\Delta \mathrm{P}}$ exhibits a noticeable dip during grokking (resulting in a slight increase in the training loss). As discussed below, this indicates that the network is undergoing a phase transition from a high entropy interconnected state to a sparse network as the system "temperature" is lowered by the optimizer. While a related quantity, namely the variance of the test loss over the sample set was previously demonstrated to reach a maximum during grokking [16,17], the dip in $\Delta$ appears to be a universal physical behavior, analogous to the maximum in the specific heat at a phase transition, as it is present in all the systems that experience grokking presented in this paper. For example, in Figure 20, the difference in test probabilities averaged over three calculations is graphed for different values of the inverse temperature for a modified **softmax** layer as well as for different values of the initialization amplitudes for a normal **softmax** layer. Evidently, as the temperature is increased, the minimum in the test amplitude occurs after fewer epochs, consistent with the smaller observed grokking delay, while decreasing the initialization amplitude first shifts the phase transition to smaller delays and subsequently eliminates it entirely as the amplitude minimum vanishes. Furthermore, as expected, the variance of $\Delta P$ increases as its mean value decreases and analogous quantities formed from the output "probabilities" are expected to display similar features.

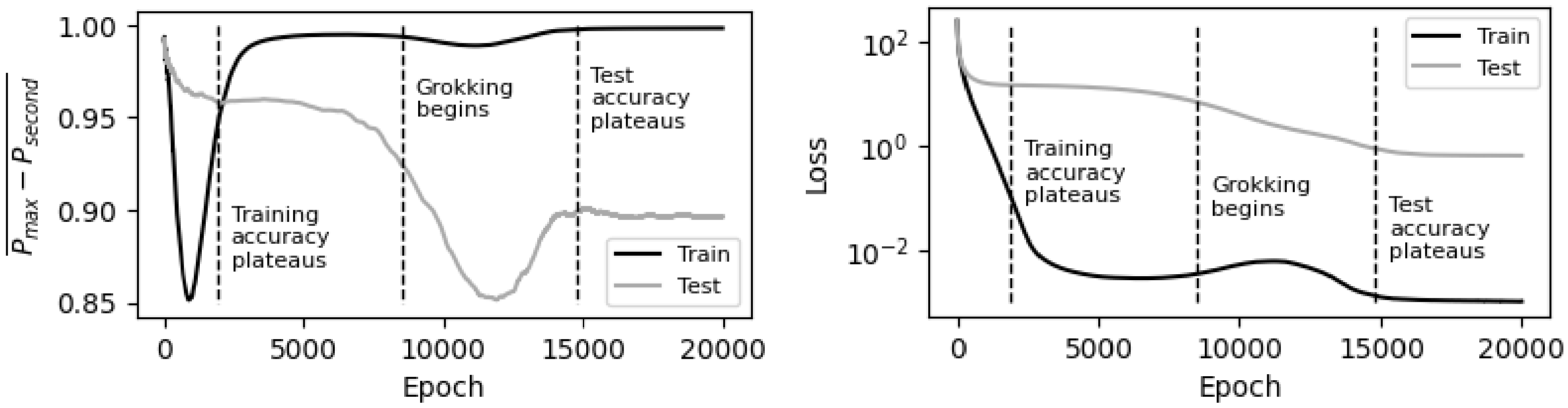

*Figure 19. (a) The average probability difference as a function of epoch number with default hyperparameters. (b) The corresponding loss curves.*

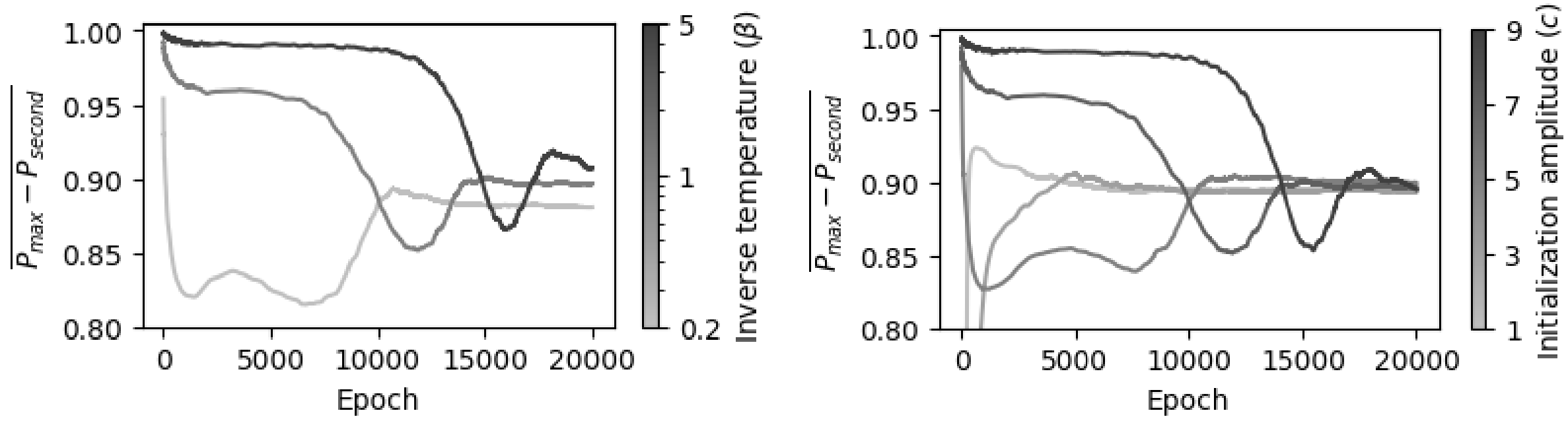

*Figure 20. The average probability difference for the test dataset for (a) varying inverse temperatures in the modified* ***softmax*** *function, and (b) initialization amplitudes for the conventional* ***softmax****.*

**Discussion and Conclusions**

To our knowledge, this paper is the first to investigate grokking behavior in an Ising model framework. While grokking often results from suboptimal choices of model hyperparameters, the analysis of this paper provides insights into the functioning of neural networks in the context of spin systems. For example, by applying novel strategies for visualizing the evolution of the network properties based on the lowest-order PCA representation of layer variables, the relationship between the test accuracy and both the network structure and the discrimination ability of each layer were ascertained. Evidently, to determine precisely the properties of previously unseen input data, the network must be sufficiently sparse. The probability that the data is then routed through ancillary paths that result in an incorrect final classification is then small. During generalization, the optimizer evaluates gradients in a wide range of directions. This increases the network sparsity while leaving the training accuracy largely unaffected. The novel PCA layer studies of the weights and gradients in this paper quantify the evolution during this process toward a network architecture that separates the test classes.

The results presented above further suggest a heuristic analogy between grokking and phase transitions. Associating an effective "entropy" such as the weight distribution entropy [18] with the network, then in the absence of dropout, grokking is accompanied by a "phase transition" from a highly interconnected and hence large entropy "liquid" network state to a "solid" stable low entropy sparse network with a minimum value of the cost function. This transition proceeds through the action of the optimizer on the network, which can be envisioned as a deterministic version of simulated annealing that decreases the network "temperature" and hence entropy through contact with a computational heat bath that absorbs entropy. This entropy, which is close to zero before and after grokking but becomes large during grokking, was characterized here through relevant features of the output layer probability distribution. Increasing the temperature associated with the modified **softmax** function in this layer increases the entropy at the output layer which promotes system reconfiguration and therefore the decreases the grokking delay (phase transition time). Additionally, dropout introduces an additional "thermal motion" that precludes a sudden transition to a sparse low-entropy network once a sufficient amount of entropy is removed from the network. Instead, the network entropy decreases continually as the temperature is lowered, resulting in a continuous approach to a "supercooled liquid" state.

While the above heuristic model provides a conceptual framework for understanding grokking behavior, the statistical quantities involved in neural network transitions cannot be unambiguously quantified in contrast to classical statistical mechanics. Thus, for example numerous strategies exist for estimating the entropy associated with a network. However, the universal behavior associated with grokking in the Ising model possesses intrinsic significance since it elucidates the qualitative behavior of the spin system in response to a gradient optimization strategy. A comparison with grokking in other spin models could accordingly provide additional insight into the relationship between network properties and generalization, especially for physical and biological systems that can be approximated by lattice models. A more challenging area of possible future research concerns the derivation of an analytic procedure that accurately describes grokking behavior. This, however, appears difficult since both the data and the network are intrinsically complex and asymmetric.

## References

[1] A. Power, Y. Burda, H. Edwards, I. Babuschkin, and V. Misra, *Grokking: Generalization Beyond Overfitting on Small Algorithmic Datasets*, arXiv:2201.02177.

[2] N. Nanda, L. Chan, T. Lieberum, J. Smith, and J. Steinhardt, *Progress Measures for Grokking via Mechanistic Interpretability*, arXiv:2301.05217.

[3] Z. Liu, E. J. Michaud, and M. Tegmark, *Omnigrok: Grokking Beyond Algorithmic Data*, arXiv:2210.01117.

[4] S. Murty, P. Sharma, J. Andreas, and C. D. Manning, *Grokking of Hierarchical Structure in Vanilla Transformers*, arXiv:2305.18741.

[5] J. Miller, C. O'Neill, and T. Bui, *Grokking Beyond Neural Networks: An Empirical Exploration with Model Complexity*, arXiv:2310.17247.

[6] A. I. Humayun, R. Balestriero, and R. Baraniuk, *Deep Networks Always Grok and Here Is Why*, arXiv:2402.15555.

[7] Z. Liu, O. Kitouni, N. Nolte, E. J. Michaud, M. Tegmark, and M. Williams, *Towards Understanding Grokking: An Effective Theory of Representation Learning*, arXiv:2205.10343.
[8] S. Golechha, *Progress Measures for Grokking on Real-World Tasks*, arXiv:2405.12755.
[9] J. Frankle and M. Carbin, *The Lottery Ticket Hypothesis: Finding Sparse, Trainable Neural Networks*, arXiv:1803.03635.
[10] D. Yunis, K. K. Patel, S. Wheeler, P. Savarese, G. Vardi, K. Livescu, M. Maire, and M. R. Walter, *Approaching Deep Learning through the Spectral Dynamics of Weights*, arXiv:2408.11804.
[11] W. Merrill, N. Tsilivis, and A. Shukla, *A Tale of Two Circuits: Grokking as Competition of Sparse and Dense Subnetworks*, arXiv:2303.11873.
[12] Z. Li et al., *The Lazy Neuron Phenomenon: On Emergence of Activation Sparsity in Transformers*, arXiv:2210.06313.
[13] G. Gur-Ari, D. A. Roberts, and E. Dyer, *Gradient Descent Happens in a Tiny Subspace*, arXiv:1812.04754.
[14] Y. Wu, T. Li, X. Cheng, J. Yang, and X. Huang, *Low-Dimensional Gradient Helps Out-of-Distribution Detection*, arXiv:2310.17163.
[15] H. Xuan, B. Yang, and X. Li, *Exploring the Impact of Temperature Scaling in Softmax for Classification and Adversarial Robustness*, arXiv:2502.20604.
[16] A. Salah and D. Yevick, *Tracing the Path to Grokking: Embeddings, Dropout, and Network Activation*, arXiv:2507.11645.
[17] A. Salah and D. Yevick, *Controlling Grokking with Nonlinearity and Data Symmetry*, arXiv:2411.05353.
[18] A. de Mathelin, F. Deheeger, M. Mougeot, and N. Vayatis, *Deep Out-of-Distribution Uncertainty Quantification via Weight Entropy Maximization*, 2025 *J. Mach. Learn*. Res. **26** 1.